\documentclass[10pt,journal]{IEEEtran}
\IEEEoverridecommandlockouts
\usepackage{mathrsfs}
\usepackage{amssymb}
\usepackage{mathbbold}
\usepackage{cite}
\usepackage{color}
\ifCLASSINFOpdf
  \usepackage[pdftex]{graphicx}
     \graphicspath{{../pdf/}{../jpeg/}}
    \DeclareGraphicsExtensions{.pdf,.jpeg,.png}
\else
   \usepackage[dvips]{graphicx}
   \graphicspath{{../eps/}}
   \DeclareGraphicsExtensions{.eps}
\fi
\usepackage[cmex10]{amsmath}
\usepackage{algorithmic}
\usepackage{array}
\usepackage{bm}
\usepackage{eqparbox}
\usepackage{mdwmath}
\usepackage{mdwtab}
\usepackage[tight,footnotesize]{subfigure}
\usepackage[caption=1]{caption}
\usepackage[font=footnotesize]{subfig}
\usepackage{fixltx2e}

\usepackage{url}

\begin{document}
\title {Sum Rate Optimization for Two Way Communications with Intelligent Reflecting Surface}
\author{Yu Zhang, Caijun Zhong,~\IEEEmembership{Senior Member,~IEEE,} Zhaoyang Zhang,~\IEEEmembership{Senior Member,~IEEE,}~and Weidang Lu
\thanks{
Y. Zhang (corresponding author) and W. Lu are with the College of Information Engineering, Zhejiang University of Technology, China. (e-mail:~{\tt yzhang@zjut.edu.cn},~{\tt luweid@zjut.edu.cn}). Y. Zhang is also with National Mobile Communications Research Laboratory, Southeast University, China.

C. Zhong and Z. Zhang are with the Department of Information Science and Electronic Engineering, Zhejiang University, China. (e-mail:~{\tt caijunzhong@zju.edu.cn},~{\tt ning\_ming@zju.edu.cn}).
}}
\maketitle

\begin{abstract}
In this letter, an intelligent reflecting surface (IRS) enhanced full-duplex MIMO two-way communication system is studied. The system sum rate is maximized through jointly optimizing the source precoders and the IRS phase shift matrix. Adopting the idea of Arimoto-Blahut algorithm, the non-convex optimization problem is decoupled into three sub-problems, which are solved alternatingly. All the sub-problems can be solved efficiently with closed-form solutions. In addition, practical IRS assumptions, e.g., discrete phase shift levels, are also considered. Numerical results verify the convergence and performance of the proposed scheme.
\end{abstract}
\section{Introduction}
Intelligent Reflecting Surface (IRS) has recently emerged as a promising technique to improve the performance of communication links \cite{Liaskos2018,Wu2019CSI}. In particular, the IRS is composed of a large number of electromagnetically reconfigurable reflective elements, and can be made extremely low-cost and energy efficient\cite{Renzo2019}. Therefore, it has received considerable research interests.



\textcolor{black}{Thus far, IRS has been considered to be incorporated into various wireless communications and technologies, such as the MIMO system, e.g., \cite{Guo2019,Wu2018,Wu2019dis,Nadeem2019}, simultaneous wireless information and power transfer (SWIPT)\cite{Pan2019}, index modulation\cite{Basar2019}, and non-orthogonal multiple access (NOMA)\cite{Fu2019}, etc. \textcolor{black}{Multi-IRS scenario has been investigated in \cite{ZLI2019}.} Considerable performance gain has been shown from the IRS assistance. Nevertheless, all the above works focused on one-way communications and to the best of our knowledge, IRS-aided multi-antenna two-way communications has not been considered yet. It is worth noting that, deploying IRS to enhance two-way communications has appealing advantages and differs from the existing related technologies such as two-way amplify-and-forward (AF) relaying \cite{Wang2010,Zhang2016}. Explicitly, since IRS only reflects the RF signals, it requires no transmit power consumption, and the issues of rate loss in half-duplex relaying and self-loop interference cancelation in full-duplex relaying do not exist in the case of IRS.}



Motivated by the above, this letter considers an IRS enhanced full-duplex MIMO bidirectional communication system, and pursues a detailed study on the joint design of the source precoders and IRS phase shift matrix maximizing the sum rate of the system. To tackle the resultant non-convex optimization problem, we exploit the structure of Arimoto-Blahut algorithm\cite{Blahut1972}, which has been adopted in the MIMO broadcast system\cite{Wang2010TWC} and two way MIMO relay system\cite{Wang2010}. Based on this, we propose an alternating approach to find a suboptimal solution. Furthermore, practical IRS restrictions, e.g., discrete phase shift levels, are also discussed. Simulation results show that the proposed algorithm achieves superior performance compared with the heuristic benchmark schemes.

\textit{Notation}: For matrices $\bf A$ and $\bf B$, $|{\bf A}|$, $tr\left({\bf A}\right)$, ${\bf A}^*$ and ${\bf A}^H$ denote the determinant, trace, conjugate, and conjugate transpose of $\bf A$. ${\bf A}\odot{\bf B}$ denotes the Hadamard product of $\bf A$ and $\bf B$. ${\mathbb E}\left[.\right]$ stands for the expectation. ${\bf I}_N$ denotes the $N$-by-$N$ identity matrix.

\section{System Model}
\begin{figure}[t]
\centering
\includegraphics[width=0.4\textwidth]{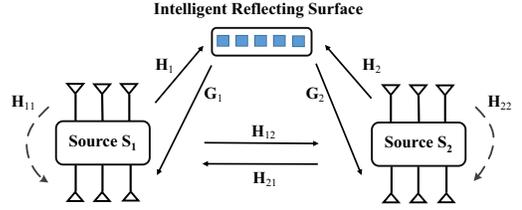}
\caption{\textcolor{black}{IRS enhanced full-duplex MIMO two-way communication systems.}} \label{fig_sys}
\end{figure}

Consider an IRS aided full-duplex MIMO bidirectional communication system as depicted in Fig. \ref{fig_sys}, which consists of two sources both equipped with $N$ transmit antennas and $N$ receive antennas, and one IRS with $M$ reflection elements. Both sources transmit to each other simultaneously with the aid of the IRS.
The transmit signal from the source $S_{i}$, $i=1,2$, is given by:
\begin{equation}
{\textbf x}_i={\bf{F}}_i{\textbf s}_i,
\end{equation}
where ${\textbf s}_i$ is the data symbol vector with unit covariance matrix ${\bf I}_N$ and ${\bf F}_i$ is the $N$-by-$N$ source precoder subject to the power constraint $tr\left({\bf F}_i{\bf F}_i^H\right) \le P$. \textcolor{black}{The IRS re-scatters the superposition of all incident signals \cite{Guo2019,Wu2018}.} \textcolor{black}{Assuming only first-order reflection from IRS \cite{Wu2018}, the reflected signal is:}
\begin{equation}
{\textbf x}_R={\bf \Theta}\left({\bf H}_1{\bf x}_1+{\bf H}_2{\bf x}_2\right),
\end{equation}
where ${\bf H}_i, i=1,2,$ denotes the $M$-by-$N$ channel matrix from the source $S_i$ to the IRS, and ${\bf \Theta}=\sqrt\eta{ diag}\left(\theta_1, \theta_2, ..., \theta_M \right)$ is the phase shift matrix of IRS, where $\eta \le 1$ is the reflection efficiency, $\theta_m, m=1,...,M$, is the reflection coefficient of the $m$th IRS element. Considering the practical implementation of IRS, three cases for the feasibility set of $\theta_m$ are assumed \cite{Guo2019}:

\begin{enumerate}
\item Each IRS element can continuously control both the amplitude and phase of the reflected signal, i.e., $\theta_m \in \mathbb{F}_1  \triangleq \left\{\theta_m\left|{\left|\theta_m\right|^2 \le 1}\right. \right\}$.
\item Each element can only adjust the phase, i.e., $\theta_m \in \mathbb{F}_2  \triangleq \left\{\theta_m\left|{\left|\theta_m\right|^2 = 1 }\right.\right\}$.
\item Each element can only take finite phase shift levels. Assume that there are $\tau$ levels equally spaced within $\left[0,2\pi \right)$, then $\theta_m \in \mathbb{F}_3  \triangleq \left\{\theta_m\left|{\theta_m = e^{j\phi_m},\phi_m\in\left\{ 0,\frac{{2\pi }}{\tau },...,\frac{{\left( {\tau  - 1} \right)2\pi }}{\tau }\right\}}\right. \right\}$.
\end{enumerate}

\textcolor{black}{Each source receives the reflected signal from the IRS as well as the signal from the other side.} For the source $S_i, i=1,2$, the received signal is:
\begin{equation}\label{rec_sig_unremove}
\begin{split}
\textcolor{black}{{{\bf{y}}_i}} & \textcolor{black}{= {{\bf{G}}_i} {{\bf{x}}_R} + {\bf H}_{{\bar i}i}{\bf x}_{\bar i}+ {{\bf{{ H}}}_{ii}}{\bf x}_i + {{\bf z}_i}}\\
 &\textcolor{black}{={\bf\Phi}_{i}{{\bf{F}}_{\bar i}}{{\bf{s}}_{\bar i}} + {{\bf{G}}_i}{\bf\Theta} {{\bf{{H}}}_i}{{\bf{F}}_i}{{\bf{s}}_i} + {{\bf{{ H}}}_{ii}}{{\bf{F}}_i}{{\bf{s}}_i} + {{\bf z}_i},}
\end{split}
\end{equation}
where ${\bf\Phi}_{i}\triangleq{{\bf G}_{i}}{\bf\Theta}{\bf H}_{\bar i}+{\bf H}_{{\bar i}i}$, ${\bar i}\triangleq3-i$, ${\bf{G}}_i$ is the $N$-by-$M$ channel matrix from the IRS to the source $S_i$, \textcolor{black}{${\bf{H}}_{{\bar i}i}$ is the $N$-by-$N$ channel matrix from the source $S_{\bar i}$ to the source $S_i$,} ${\bf H}_{ii}$ is the $N$-by-$N$ residual self-loop interference matrix \cite{Zhang2016} at the source $S_i$ (due to the non-ideal full-duplex signal processing), and ${{\bf z}_i}$ is the additive white Gaussian noise with normalized covariance ${\bf I}_N$. Assume that the channel state information (CSI) of each link as well as $\bf\Theta$ is perfectly known by each source, (CSI acquisition has been discussed in, e.g., \cite{Wu2019CSI}). \textcolor{black}{Then the source $S_i$ can subtract the term ${{\bf{G}}_i}{\bf\Theta} {{\bf{{H}}}_i}{{\bf{F}}_i}{{\bf{s}}_i}$ from its received signal \eqref{rec_sig_unremove}\footnote{\textcolor{black}{Note that $S_i$ is aware of its own data symbol ${\bf{s}}_i$ and precoder ${\bf{F}}_i$.}}.} We have:
\begin{equation}\label{rec_sig}\textcolor{black}{
{{\bf{y}}_i} = {\bf\Phi}_i{{\bf{F}}_{\bar i}}{{\bf{s}}_{\bar i}}  + {{\bf{{ H}}}_{ii}}{{\bf{F}}_i}{{\bf{s}}_i} + {{\bf z}_i}.}
\end{equation}

From \eqref{rec_sig}, the achievable rate for source $S_i$, $i=1,2$, is given by:
\begin{equation}\label{achivable_rate}
\textcolor{black}{{R_i}= \log \left| {{{\bf{I}}_M} + {\bf{F}}_i^H{\bf\Phi}_{\bar i}^H{\bf{\Omega }}_{\bar i}^{ - 1}{\bf\Phi}_{\bar i}{{\bf{F}}_i}} \right|,}
\end{equation}
where ${\bf{\Omega }}_i^{} \triangleq{{\bf{{ H}}}_{ii}}{{\bf{F}}_i}{\bf{F}}_i^H{\bf{{ H}}}_{ii}^H + {{\bf{I}}_N}$.

\section{Achievable Sum Rate Maximization}
We optimize the source precoders ${\bf F}_i, i=1,2$, and the IRS phase shift matrix $\bf \Theta$ to maximize the system sum rate. The optimization problem is formulated as follows:
\begin{equation}\label{oring_problem}
\begin{split}
\mathop {\max }\limits_{{{\bf{F}}_1},{{\bf{F}}_2},{\bf\Theta} }~~& \sum\limits_{i = 1}^2 {{R_i}} \\
s.t.~~&
tr\left( {{{\bf{F}}_i}{\bf{F}}_i^H} \right) \le P,~~i=1,2\\
&{\theta _m} \in {\mathbb F},~~m = 1,...,M,
\end{split}
\end{equation}
where ${\mathbb F}$ can be ${\mathbb F}_1$, ${\mathbb F}_2$ or ${\mathbb F}_3$ which are defined in the previous section.
Unfortunately, it is not straightforward to solve  \eqref{oring_problem} even for ${\mathbb F}={\mathbb F}_1$ due to its non-convexity. Here we adopt the Arimoto-Blahut structure\cite{Blahut1972}. Before the derivation, we introduce the following lemma \textcolor{black}{[16, Lemma 10.8.1, p. 333]}:

\newtheorem{lemma}{Lemma}
\begin{lemma}\label{basic_lemma}
For a channel with input $s$, output $y$ and the transition probability $p\left(y\left|s\right.\right)$, the mutual information $I\left(s;y\right)$ with an arbitrary input probability distribution $p\left(s\right)$ is given by:
\begin{equation}
I\left( {s;y} \right) = \mathop {\max }\limits_{q(s|y)} {\mathbb E}\left[ {\log \left( {\frac{{q\left( {s|y} \right)}}{{p\left( s \right)}}} \right)} \right],
\end{equation}
where the expectation is taken over all possible $s$ and $y$ generated from the probability distribution $p\left(s\right)$ and $p\left(y\left|s\right.\right)$. The optimal $q^o\left(s|y\right)$ is the posterior probability:
\begin{equation}
{q^o}\left( {s|y} \right) = \frac{{p\left( s \right)p\left( {y|s} \right)}}{{p\left( y \right)}} \triangleq p\left( {s|y} \right).
\end{equation}
\end{lemma}

Note that the achievable rate \eqref{achivable_rate} of the source $S_i$ is derived from $I\left({\bf s}_i; {\bf y}_{\bar i}\right)$ where the input probability distribution $p\left({\bf s}_{ i}\right)$ is ${\mathcal {CN}}\left({\bf 0},{\bf I}_N\right)$ and the channel transition probability $p({{\bf y}_{\bar i}}|{{\bf s}_i})$ is from \eqref{rec_sig}. Then according to Lemma \ref{basic_lemma}, \eqref{achivable_rate} can also be re-expressed as:
\begin{equation}\label{R_anotherform}
{R_i} = \mathop {\max }\limits_{q({{\bf s}_i}|{{\bf y}_{\bar i}})} {\mathbb E}\left[ {\log \left( {\frac{{q\left( {{{\bf s}_i}|{{\bf y}_{\bar i}}} \right)}}{{\mathcal {CN}\left( {{\bf 0},{\bf{I}}_N} \right)}}} \right)} \right], i=1,2.
\end{equation}
The optimal $q^o({{\bf s}_i}|{{\bf y}_{\bar i}})$ is the posterior probability $p\left({{\bf s}_i|{\bf y}_{\bar i}} \right)$. According to \textcolor{black}{[17, Theorem 10.3, p. 326]}, it can be derived that $p\left({{\bf s}_i|{\bf y}_{\bar i}} \right)$ follows the complex Gaussian distribution $\mathcal {CN}\left( {{{\bf{W}}^o_{\bar i}}{{\bf y}_{\bar i}},{{\bf{\Sigma }}^o_{\bar i}}} \right)$ with:
\begin{equation}\label{opt_W}
{{\bf{W}}^o_{\bar i}} = {\bf U}_{\bar i}^H{\left( {{{\bf U}_{\bar i}}{\bf U}_{\bar i}^H + {{\bf \Omega}_{\bar i}}} \right)^{ - 1}},
\end{equation}
\begin{equation}\label{opt_Sigma}
{{\bf{\Sigma }}^o_{\bar i}} = {{\bf I}_N} - {{\bf{W}}^o_{\bar i}}{{\bf U}_{\bar i}},
\end{equation}
where \textcolor{black}{${{\bf U}_{\bar i}} \triangleq \left({{\bf{G}}_{\bar i}}{\bf\Theta} {{\bf{{ H}}}_{ i}}+{\bf H}_{i\bar i}\right){{\bf{F}}_{ i}}$}. According to \eqref{R_anotherform}$-$\eqref{opt_Sigma}, the problem \eqref{oring_problem} can be re-formulated as follows:
\begin{equation}\label{new_problem}
\begin{split}
\mathop {\max }\limits_{{{\bf{F}}_1},{{\bf{F}}_2},{\bf \Theta} ,{{\bf{W}}_1},{{\bf{\Sigma }}_1},{{\bf{W}}_2},{{\bf{\Sigma }}_2}} & \sum\limits_{i = 1}^2 {{\mathbb E}\left[ {\log \left( {\frac{{\mathcal {CN}\left( {{{\bf{W}}_{\bar i}}{{\bf y}_{\bar i}},{{\bf{\Sigma }}_{\bar i}}} \right)}}{{\mathcal {CN}\left( {{\bf 0},{\bf{I}}_N} \right)}}} \right)} \right]} \\
s.t.~~&
tr\left( {{{\bf{F}}_i}{\bf{F}}_i^H} \right) \le P,~~i=1,2\\
&{\theta _m} \in {\mathbb F},~~m = 1,...,M.
\end{split}
\end{equation}

To this end, we tackle the above problem using the alternating optimization approach by iteratively solving three subproblems.
\subsection{Update ${{\bf{W}}_1},{{\bf{\Sigma }}_1},{{\bf{W}}_2}$, and ${{\bf{\Sigma }}_2}$}
We optimize ${{\bf{W}}_i}$ and ${{\bf{\Sigma }}_i}$ under fixed ${\bf F}_i$ and ${\bf\Theta}$, $i=1,2$. From \eqref{new_problem}, the problem is written as:
\begin{equation}\label{sub_A}
\begin{split}
\mathop {\max }\limits_{{{\bf{W}}_1},{{\bf{\Sigma }}_1},{{\bf{W}}_2},{{\bf{\Sigma }}_2}} & \sum\limits_{i = 1}^2 {{\mathbb E}\left[ {\log \left( {\frac{{\mathcal {CN}\left( {{{\bf{W}}_{\bar i}}{{\bf y}_{\bar i}},{{\bf{\Sigma }}_{\bar i}}} \right)}}{{\mathcal {CN}\left( {{\bf 0},{\bf{I}}_N} \right)}}} \right)} \right]}.
\end{split}
\end{equation}
Obviously, the solution has already been given by \eqref{opt_W} and \eqref{opt_Sigma}.
\subsection{Update the IRS phase shift matrix $\bf\Theta$}
We optimize ${\bf\Theta}$ under fixed ${\bf F}_i$, ${{\bf{W}}_i}$ and ${{\bf{\Sigma }}_i}$, $i=1,2$. Firstly, we calculate the expectation term in the objective function of \eqref{new_problem} as follows:
\begin{equation}\label{Eq_expend_expec}
\begin{split}
&-{\mathbb E}\left[ {\log \left( \frac{{\mathcal {CN}\left( {{{\bf{W}}_{\bar i}}{{\bf y}_{\bar i}},{{\bf{\Sigma }}_{\bar i}}} \right)}}{{{\mathcal {CN}\left( {{\bf 0},{{\bf{I}}_N}} \right)}}} \right)} \right]\\
=&~{\mathbb E}\left[ {{{\left( {{\bf x}_i - {{\bf{W}}_{\bar i}}{{\bf y}_{\bar i}}} \right)}^H}{\bf \Sigma} _{\bar i}^{ - 1}\left( {{\bf x}_i - {{\bf{W}}_{\bar i}}{{\bf y}_{\bar i}}} \right)} \right] + \log \left| {{{\bf \Sigma} _{\bar i}}} \right| + N\log \pi\\
&- {\mathbb E}\left[ {{{\bf x}_i}{\bf x}_i^H} \right] - \log \left| {{{\bf{I}}_N}} \right| - N\log \pi \\
=&~\textcolor{black}{tr\left( {{\bf{W}}_{\bar i}^H{\bf \Sigma} _{\bar i}^{ - 1}{{\bf{W}}_{\bar i}}{\bf\Phi}_{\bar i}{{\bf{F}}_{ i}}{\bf{F}}_{i}^H{\bf\Phi}_{\bar i}^H} \right)+tr\left( {{\bf{W}}_{\bar i}^H{\bf \Sigma} _{\bar i}^{ - 1}{{\bf{W}}_{\bar i}}{{\bf\Omega} _{\bar i}}} \right)} \\
&~\textcolor{black}{
 - 2{\mathop{\rm Re}\nolimits} \left( tr\left( {{\bf\Sigma} _{\bar i}^{ - 1}{{\bf{W}}_{\bar i}}{\bf\Phi}_{\bar i}{{\bf{F}}_{i}}} \right)\right)+tr\left( {{\bf \Sigma} _{\bar i}^{ - 1}} \right) + \log \left| {{{\bf\Sigma} _{\bar i}}} \right| - N. }\\
\end{split}
\end{equation}

In the last equality of \eqref{Eq_expend_expec}, the first and third terms include the phase shift matrix $\bf\Theta$. Let ${\bf A}_i$ and ${\bf B}_i$ denote the terms ${\bf{G}}_{\bar i}^H{\bf{W}}_{\bar i}^H{\bf \Sigma} _{\bar i}^{ - 1}{{\bf{W}}_{\bar i}}{{\bf{G}}_{\bar i}}$ and $\left({{\bf{{ H}}}_{ i}}{{\bf{F}}_{ i}}{\bf{F}}_{ i}^H{\bf{{ H}}}_{ i}^H\right)^T$, respectively. The first term in \eqref{Eq_expend_expec} can be written as:
\begin{equation}\label{def_AB}
\begin{split}
&\textcolor{black}{tr\left( {{\bf{W}}_{\bar i}^H{\bf \Sigma} _{\bar i}^{ - 1}{{\bf{W}}_{\bar i}}{\bf\Phi}_{\bar i}{{\bf{F}}_{ i}}{\bf{F}}_{i}^H{\bf\Phi}_{\bar i}^H} \right)}\\
=&~tr\left( {{{\bf\Theta} ^H}{\bf{G}}_{\bar i}^H{\bf{W}}_{\bar i}^H{\bf \Sigma} _{\bar i}^{ - 1}{{\bf{W}}_{\bar i}}{{\bf{G}}_i}{\mathbf \Theta} {{\bf{{ H}}}_{ i}}{{\bf{F}}_{ i}}{\bf{F}}_{ i}^H{\bf{{ H}}}_{ i}^H} \right)\\
&~\textcolor{black}{+2{\rm Re}\left(tr\left({\bf H}_i{\bf F}_i{\bf F}_i^H{\bf H}_{i{\bar i}}^H{\bf{W}}_{\bar i}^H {\bf \Sigma} _{\bar i}^{ - 1}{{\bf{W}}_{\bar i}}{{\bf{G}}_{\bar i}}{\bf \Theta}  \right) \right)} \\
&~+tr\left({\bf{W}}_{\bar i}^H {\bf \Sigma} _{\bar i}^{ - 1}{{\bf{W}}_{\bar i}}{\bf H}_{i{\bar i}}{\bf F}_{i}{\bf F}_{i}^H{\bf H}^H_{i{\bar i}} \right)\\
\mathop  = \limits^{(a)}&~\eta{{\bm\theta} ^H}\left( {{{\bf A}_i} \odot {{\bf B}_i}} \right){\bm\theta}+tr\left({\bf{W}}_{\bar i}^H {\bf \Sigma} _{\bar i}^{ - 1}{{\bf{W}}_{\bar i}}{\bf H}_{i{\bar i}}{\bf F}_{i}{\bf F}_{i}^H{\bf H}^H_{i{\bar i}} \right)\\
&~~\textcolor{black}{+2\sqrt{\eta}{\rm Re}\left({\bm \theta}^H{\bf{d}}_i^*\right)},
\end{split}
\end{equation}
\textcolor{black}{where ${\bm\theta}\triangleq\left[\theta_1,\theta_2,...,\theta_M\right]^T$, ${\bf{d}}_i$ denotes a $M$-by-$1$ vector which consists of the diagonal entries of the matrix ${\bf H}_i{\bf F}_i{\bf F}_i^H{\bf H}_{i{\bar i}}^H{\bf{W}}_{\bar i}^H{\bf \Sigma} _{\bar i}^{ - 1}{{\bf{W}}_{\bar i}}{{\bf{G}}_{\bar i}}$ and $(a)$ is due to $tr\left({\bf\Theta}^H{\bf A}_i{\bf\Theta}{\bf B}_i^T\right)=\eta{{\bm\theta} ^H}\left( {{{\bf A}_i} \odot {{\bf B}_i}} \right){\bm\theta}$.}
For the third term in \eqref{Eq_expend_expec}:
\begin{equation}\label{def_C}
\begin{split}
&tr\left( {{\bf\Sigma} _{\bar i}^{ - 1}{{\bf{W}}_{\bar i}}{\bf\Phi}_{\bar i}{{\bf{F}}_{ i}}} \right)\\
=&~\textcolor{black}{tr\left( {{\bf{{ H}}}_{ i}}{{\bf{F}}_{ i}}{{\bf\Sigma} _{\bar i}^{ - 1}{{\bf{W}}_{\bar i}}{{\bf{G}}_{\bar i}}{\bf\Theta} } \right)
+tr\left({\bf\Sigma} _{\bar i}^{ - 1}{{\bf{W}}_{\bar i}}{\bf H}_{i{\bar i}}{\bf F}_{i} \right)}\\
=&~\textcolor{black}{\sqrt{\eta}\left({\bm \theta}^H{\bf{b}}_i^*\right)^*+tr\left({\bf\Sigma} _{\bar i}^{ - 1}{{\bf{W}}_{\bar i}}{\bf H}_{i{\bar i}}{\bf F}_{i} \right)},
\end{split}
\end{equation}
\textcolor{black}{where ${\bf{b}}_i$ denotes a $M$-by-$1$ vector which is composed of the diagonal entries of the product ${{\bf{{ H}}}_{ i}}{{\bf{F}}_{ i}}{\bf\Sigma} _{\bar i}^{ - 1}{{\bf{W}}_{\bar i}}{{\bf{G}}_{\bar i}}$.}

Substitute \eqref{Eq_expend_expec}-\eqref{def_C} into the objective function of the problem \eqref{new_problem} and remove the terms irrelevant to $\bf\Theta$, the sub-problem optimizing $\bf\Theta$ is given by:
\begin{equation}\label{problem_theta}
\begin{split}
\mathop {\max }\limits_{\bm\theta}&- {{\bm\theta} ^H}\left( \sum\limits_{i = 1}^2 {\eta {{\bf A}_i} \odot {{\bf B}_i}} \right){\bm\theta}  + 2{\mathop{\rm Re}\nolimits} \left( {{{\bm\theta} ^H}\sum\limits_{i = 1}^2 {\sqrt \eta  {\bf c}_i^*}} \right)\\
s.t.~~&{\theta _m} \in {\mathbb F},~~m = 1,...,M,
\end{split}
\end{equation}
\textcolor{black}{where ${\bf c}_i\triangleq {\bf b}_i-{\bf d}_i$.} We first consider the case ${\mathbb F}={\mathbb F}_1$. The cases for ${\mathbb F}_2$ and ${\mathbb F}_3$ will be discussed later. Now the constraint in \eqref{problem_theta} is ${\left|\theta_m\right|^2 \le 1}$ which is convex on $\theta_m$. We rewrite this constraint in a quadratic form as follows:
\begin{equation}
{{\bm\theta} ^H}{{\bm\varepsilon} _m}{\bm\varepsilon} _m^H{\bm\theta}  \le 1,~~m = 1,...,M,
\end{equation}
where ${\bm\varepsilon}_m$ denotes an $M$-by-$1$ vector whose elements are all zero except that the $m$th is one. It is easy to verify that ${\bf A}_i$ and ${\bf B}_i$, $i=1,2$, are Hermitian semi-positive definite matrices. Then the Hadamard product ${\bf A}_i\odot{\bf B}_i$ is also semi-positive definite. Therefore, \eqref{problem_theta} with ${\mathbb F}={\mathbb F}_1$ is a convex Quadratic Programming with Quadratic Constraints (QCQP), which can be solved efficiently through Lagrange dual method \cite{Boyd2004}. The solution of \eqref{problem_theta} is given by:
\begin{equation}
{{\bm\theta} ^o} = \left( \sum\limits_{m = 1}^M {{\lambda_m^o}{{\bm\varepsilon}_m}{\bm\varepsilon} _m^H}  + \sum\limits_{i = 1}^2 {\eta {{\bf A}_i} \odot {{\bf B}_i}} \right)^{-1}\sum\limits_{i = 1}^2 {\sqrt \eta  {\bf c}_i^*},
\end{equation}
where $\lambda_m^o$, $m=1,...,M$, are the optimal Lagrange dual variables, which can be obtained through the sub-gradient method or ellipsoid method\cite{Boyd2004}.
\subsection{Update the source precoders ${\bf F}_1$ and ${\bf F}_2$}
We optimize ${\bf F}_1$ and ${\bf F}_2$ when ${{\bf{W}}_1},{{\bf{\Sigma }}_1},{{\bf{W}}_2}$, ${{\bf{\Sigma }}_2}$ and $\bm\Theta$ hold fixed. Recalling the last equality in \eqref{Eq_expend_expec}, it can be found that the first three terms are related to ${\bf F}_{1}$ and ${\bf F}_{2}$. Therefore, the optimization problem is given by:
\begin{equation}\label{Problem_F}
\begin{split}
\mathop {\max }\limits_{{{\bf{F}}_1},{{\bf{F}}_2}}  &\textcolor{black}{- \sum\limits_{i = 1}^2 {tr\left( {{\bf{W}}_{\bar i}^H{\bf \Sigma} _{\bar i}^{ - 1}{{\bf{W}}_{\bar i}}{\bf\Phi}_{\bar i}{{\bf{F}}_{ i}}{\bf{F}}_{i}^H{\bf\Phi}_{\bar i}^H} \right)}}\\
&~~ - \sum\limits_{i = 1}^2 {tr\left( {{\bf{W}}_{\bar i}^H{\bf\Sigma} _{\bar i}^{ - 1}{{\bf{W}}_{\bar i}}{{\bf{{ H}}}_{\bar i\bar i}}{{\bf{F}}_{\bar i}}{\bf{F}}_{\bar i}^H{\bf{{ H}}}_{\bar i\bar i}^H} \right)}  \\
&~~\textcolor{black}{+ 2\sum\limits_{i = 1}^2 {{\mathop{\rm Re}\nolimits} \left(tr\left( {{\bf\Sigma} _{\bar i}^{ - 1}{{\bf{W}}_{\bar i}}{\bf\Phi}_{\bar i}{{\bf{F}}_i}}\right) \right)}}\\
s.t.&~~tr\left( {{{\bf{F}}_i}{\bf{F}}_i^H} \right) \le P,~~i=1,2.
\end{split}
\end{equation}
Obviously, in the above problem, ${\bf F}_1$ or ${\bf F}_2$ can be optimized individually. We divide the problem into two sub-problems, each of which has only ${\bf F}_1$ or ${\bf F}_2$ as the optimizing variable. Taking ${\bf F}_1$ as an example, the problem is written as:
\begin{equation}\label{Problem_F_sub}
\begin{split}
\mathop {\max }\limits_{{{\bf{F}}_1}} & - tr\left( {{{\bf{F}}_1^H}{{\bf J}_1}{\bf{F}}_1} \right) + 2{\mathop{\rm Re}\nolimits} \left(tr\left( {{{\bf{F}}_1^H}}{{\bf K}_1}^H\right) \right)\\
s.t.&~~tr\left( {{{\bf{F}}_1}{\bf{F}}_1^H} \right) \le P,
\end{split}
\end{equation}
where we have the following notations:
\begin{equation}
\textcolor{black}{{{\bf J}_1} \triangleq {\bf\Phi}_{2}^H{\bf{W}}_2^H{\bf\Sigma} _2^{ - 1}{{\bf{W}}_2}{\bf\Phi}_{2}+ {\bf{{H}}}_{11}^H{\bf{W}}_1^H{\bf\Sigma} _1^{ - 1}{{\bf{W}}_1}{{\bf{{ H}}}_{11}},}
\end{equation}
\begin{equation}
\textcolor{black}{{{\bf K}_1}\triangleq {\bf\Sigma} _{2}^{ - 1}{{\bf{W}}_{2}}{\bf\Phi}_{2}.}
\end{equation}
It is straightforward that ${{\bf J}_1}$ is semi-positive definite. Therefore, \eqref{Problem_F_sub} is a convex QCQP. Its solution can be derived as:
\begin{equation}
{\bf{F}}_1^o = {\left( {\lambda^o  + {{\bf J}_1}} \right)^{ - 1}}{\bf K}_1^H,
\end{equation}
where $\lambda^o$ is the optimal Lagrange dual variable obtained through bi-section search method or sub-gradient method\cite{Boyd2004}. The precoder matrix ${\bf F}_2$ can be obtained in a similar fashion.
\subsection{Discussion}
We solve the original problem \eqref{new_problem} in an iterative manner. During each iteration, the three sub-problems \eqref{sub_A}, \eqref{problem_theta} and \eqref{Problem_F} are solved alternatingly. It is straightforward to see that the objective function in \eqref{new_problem} is monotonically increasing after solving each of the three sub-problems, which guarantees the convergence of the proposed optimization scheme.

For now the proposed scheme solves the original problem \eqref{new_problem} with $\theta_m\in{\mathbb F}_1$. For the other two cases ${\mathbb F}_2$ and ${\mathbb F}_3$, the constraints on $\theta_m$, $m=1,...,M$, become non-convex and the sub-problem \eqref{problem_theta} is not convex too. Here we adopt the idea in \cite{Guo2019}.  \textcolor{black}{We still apply the proposed iterative algorithm, except for the sub-problem \eqref{problem_theta}.} Explicitly, denote the solution of \eqref{problem_theta} with $\theta_m\in{\mathbb F}_1$ as $\theta_m^{o,{\mathbb F}_1}$ and its angle as $\phi_m^{o,{\mathbb F}_1}$, $m=1,...,M$. Then for the case $\theta_m\in{\mathbb F}_2$, the solution of the sub-problem is given by $\theta_m^{o,{\mathbb F}_2}=e^{j\phi_m^{o,{\mathbb F}_1}}$. For the case $\theta_m\in{\mathbb F}_3$, the solution is $\theta_m^{o,{\mathbb F}_3}=e^{j\phi_m^{o,{\mathbb F}_3}}$,
where $\phi_m^{o,{\mathbb F}_3}=\arg \mathop {\min }\limits_{\phi  \in \left\{ {0,\frac{{2\pi }}{\tau },...,\frac{{\left( {\tau  - 1} \right)2\pi }}{\tau }} \right\}} \left| {\phi _m^{o,{{\mathbb F}_1}} - \phi } \right|$. \textcolor{black}{Note that the above solution may be suboptimal. To guarantee the monotonic increase of the objective function in \eqref{new_problem}, we also check its value with the old $\bf\Theta$ in the last round. If the objective function with the new $\bf\Theta$ is smaller than that with the old one, we still use the latter as the solution of problem \eqref{problem_theta} in this round.} Note that the solution for the case $\theta_m\in{\mathbb F}_1$ is used to initialize the optimization procedure for the other two cases. Following the similar step in \cite{Wu2019dis}, it can be verified that the rate loss due to the discrete phase shift is regardless of $M$ when $N=1$ and $M$ is large.


\section{Numerical Results}
\begin{figure}[t]
\centering
\includegraphics[width=0.45\textwidth]{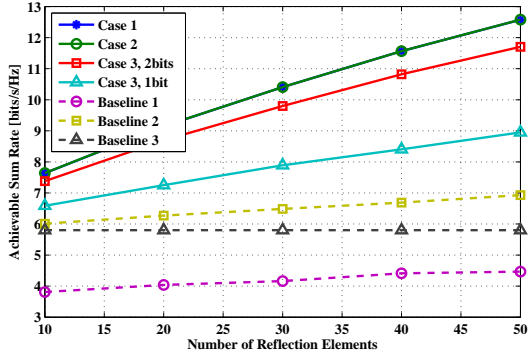}
\caption{\color{black}{Average sum rate versus number of IRS elements.}} \label{fig_rateantenna}
\end{figure}
\begin{figure}[t]
\centering
\includegraphics[width=0.45\textwidth]{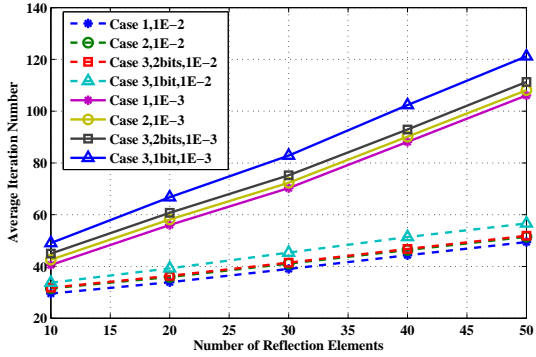}
\caption{\color{black}{Average iteration number of the proposed scheme.}} \label{fig_converg}
\end{figure}
\begin{figure}[t]
\centering
\includegraphics[width=0.45\textwidth]{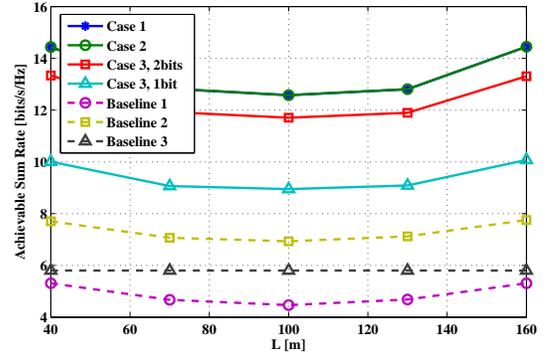}
\caption{\color{black}{Average sum rate versus the location of IRS.}} \label{fig_ratedist}
\end{figure}
\newcommand {\tabincell}[2]{\begin{tabular}{@{}#1@{}}#2\end{tabular}}
\begin{table}[]
\caption{\color{black}{Comparison with the exhaustive search scheme}}
\begin{tabular}{|p{0.55cm}|p{1cm}|p{1cm}|p{0.75cm}|p{1cm}|p{1cm}|p{0.75cm}|}
\hline
\tabincell{c}{P\\{[}dBm{]}} & \tabincell{c}{1bit\\{[}bits/s/Hz{]}} &\tabincell{c}{Ex, 1bit\\{[}bits/s/Hz{]}}&Loss&\tabincell{c} {2bits\\{[}bits/s/Hz{]}} &\tabincell{c}{Ex, 2bits\\{[}bits/s/Hz{]}} &Loss\\ \hline
8          & 1.1114                & 1.3113                    & 15.2\% & 1.3484                 & 1.4393                     & 6.3\% \\ \hline
10          & 1.5782                & 1.8730                    & 15.7\% & 1.9178                 & 2.0336                     & 5.7\% \\ \hline
12          & 2.2093                & 2.5418                    & 13.1\% & 2.5663                 & 2.7097                     & 5.3\% \\ \hline
14          & 2.8414                & 3.2837                    & 13.5\% & 3.3393                 & 3.5327                     & 5.5\% \\ \hline
16          & 3.6415                & 4.1278                    & 11.8\% & 4.1786                 & 4.4147                     & 5.3\% \\ \hline
\end{tabular}
\end{table}
\begin{figure}[t]
\centering
\includegraphics[width=0.45\textwidth]{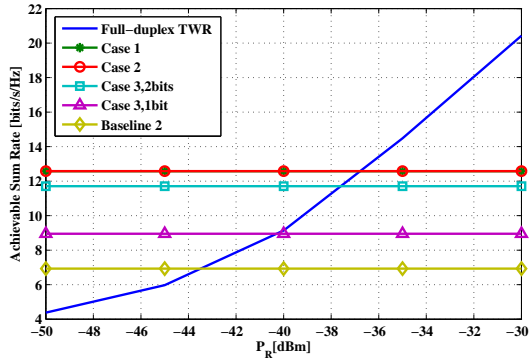}
\caption{\color{black}{Comparison with the full-duplex two way relay.}} \label{fig_raterelay}
\end{figure}

\textcolor{black}{In this section, we present numerical results to verify the performance of the proposed joint optimization scheme. In the simulation scenario, the two sources are located at $(0,0)$ and $(200,0)$, respectively. The position of the IRS is $(L,20)$. Note that the unit of distance is meter. Similar to \cite{Wu2018,Guo2019}, the large scale fading is modelled as $\kappa  = \varsigma {d^{ - \alpha }}$, where $d$ is the distance between the transmitter and receiver, $\alpha$ is the path loss exponent, and $\varsigma$ is the path loss at the reference distance $1$m which is set to $-30$dB. Since in practice the position of IRS can be carefully chosen, we assume that the IRS-source link has better channel condition compared with the direct link. For the former, the path loss exponent $\alpha$ is set to $2$ while for the latter, $\alpha=3.5$. For small-scale fading, Rayleigh fading is assumed for each link. Besides, the background noise variance at each node is set to $-110$dBm. As for the residual self-interference matrix, we assume each entry is assumed i.i.d. zero-mean complex Gaussian and the residual self-interference under $0$dBm transmit power is at the same level as the background noise. Finally, the relative reflection gain of the IRS elements over the antenna gain of both sources (assumed to be $0$dBi) is set to $5$dB \cite{Wu2018} and the reflection efficiency is set to $1$.}

Fig. \ref{fig_rateantenna} presents the average sum rate with the optimized source precoders ${\bf F}_1$, ${\bf F}_2$ and the phase shift matrix $\bf \Theta$ when $N=2$, $L=100$ and $P=5$dBm. `Case 1' and `Case 2' denote the results for $\theta_m\in{\mathbb F}_1$ and $\theta_m\in{\mathbb F}_2$, respectively. `Case 3, 2bits' represents the result for $\theta_m\in{\mathbb F}_3$ wherein the number of discrete phase levels, i.e., $\tau$, is $4$. As for `Case 3, 1bit', $\tau=2$. Three baseline schemes are also simulated. In `Baseline 1', all the three matrices are randomly generated. In `Baseline 2', $\bf\Theta$ is randomly generated while ${\bf F}_1$ and ${\bf F}_2$ are optimized. `Baseline 3' is the direct-link case without the aid of IRS, where ${\bf F}_1$ and ${\bf F}_2$ are optimized. It can be observed that the proposed scheme for ${\mathbb F}_1$, ${\mathbb F}_2$ and ${\mathbb F}_3$ all show considerably gain over the baselines. \textcolor{black}{It is also observed that the performance gap between `Case 1' and `Case 2' is negligible. It is found that in most (however, not all) channel realizations the amplitude of the optimized $\theta_m$ in `Case 1' is very close to $1$.} On the other hand, the gap between `Case 1' and `Case 3, 2bits' is considerably small, which shows that solely $4$ discrete levels of phase shift can achieve most gain.

\textcolor{black}{In Fig. \ref{fig_converg}, the average number of iterations versus IRS element number under two accuracies is given, where the simulation parameters are similar to those in Fig. \ref{fig_rateantenna}. As for the accuracy, `1E-3' means that the iteration is terminated when the gap between the value of the objective function in the current iteration and that in the previous iteration is no larger than $10^{-3}$. Note that the algorithm for `Case 2' or `Case 3' is initialized by the solution for `Case 1'. Therefore, the iteration number for the two cases should also include that for solving `Case 1'. The plotted results verify the convergence of the proposed optimization scheme for all cases.}

\textcolor{black}{We also examine the efficiency of the proposed algorithm for `Case 3', by comparing it with the exhaustive search method, wherein we exhaustively search all feasible phase shift matrix $\bf\Theta$ with the source precoders optimized correspondingly. In consideration of complexity, we solely set $N=2$ and $M=3$. Note that in this case, the signal power from the IRS will be too small compared with that from the direct link. Therefore, we assume no direct link in the simulation. In Table I, it can be observed that the loss of the `2bits' case from the exhaustive search method `Ex, 2bits' is quite small compared with that for the `1bit' case.}

\textcolor{black}{Fig. \ref{fig_ratedist} exploits the impact on the IRS location, where $P=5$dBm, $N=2$, $M=50$ and the IRS is moving from $L=40$ to $L=160$. It can be observed that the worst average sum rate is achieved when IRS is located in the middle between the two sources.
}

\textcolor{black}{Finally, in Fig.\ref{fig_raterelay}, the performance of the IRS with $M=50$ is compared with MIMO full-duplex amplify-and-forward two way relay (TWR), wherein $P=5$dBm, $N=2$. Similar to the IRS, the relay is located at $(100,20)$ and equipped with both $50$ transmit and receiver antennas. We apply the MRC/MRT  precoder \cite{Zhang2016} at the relay. It is observed that in the simulation scenario, the performance gain from IRS is comparable to that from a relay with transmit power $P_R$ solely around $-40$dBm to $-35$dBm. This is due to the fact that IRS suffers from the `double-fading' effect. However, recalling that actually IRS requires no transmit power .}
\section{Conclusions}
In this letter, we have optimized the source precoders and the IRS phase shift matrix in the full-duplex MIMO two-way communication system to maximize the system sum rate. Three cases for the phase shift at IRS have been considered. The non-convex problem has been decomposed into three sub-problems, which are solved iteratively. Simulations have verified the convergence and performance of the proposed scheme.


\begin{thebibliography}{unsrt}

\bibitem{Liaskos2018}
C.~Liaskos, S.~Nie, A.~Tsioliaridou, A.~Pitsillides, S.~Ioannidis, and
I.~Akyildiz, ``A new wireless communication paradigm through software controlled
metasurfaces," \emph{IEEE Commun. Mag.}, vol. 56, no. 9, pp. 162-169, 2018.

\bibitem{Wu2019CSI}
Q. Wu and R. Zhang, ``Towards smart and reconfigurable environment: Intelligent reflecting surface aided wireless network," \emph{arXiv preprint}
arXiv:1905.00152, 2019.

\bibitem{Renzo2019}
M. Di Renzo, M. Debbah, D.-T. Phan-Huy, A. Zappone, M.-S. Alouini,
C. Yuen, V. Sciancalepore, G. C. Alexandropoulos, J. Hoydis, and
H. Gacanin, ``Smart radio environments empowered by AI reconfigurable
meta-surfaces: An idea whose time has come," \emph{arXiv preprint}
arXiv:1903.08925, 2019.


\bibitem{Guo2019}
H. Guo, Y. C. Liang, J. Chen, and E. G. Larsson, ``Weighted Sum-Rate Optimization for Intelligent
Reflecting Surface Enhanced Wireless Networks", \emph{arXiv preprint}, arXiv:1905.07920, 2019.

\bibitem{Wu2018}
Q. Wu and R. Zhang, ``Intelligent reflecting surface enhanced wireless
network: Joint active and passive beamforming design," in Proc. \emph{IEEE
Globecom}, Dec. 2018, pp. 1-6.

\bibitem{Wu2019dis}
Q. Wu and R. Zhang, ``Beamforming Optimization for Wireless Network Aided by Intelligent Reflecting Surface with Discrete Phase Shifts," \emph{IEEE Trans. Commun.}, Dec. 2019, early access.

\bibitem{Nadeem2019}
Q.U.A. Nadeem, A. Kammoun, A. Chaaban, M. Debbah, and M.S. Alouini, ``Asymptotic analysis of large intelligent
surface assisted MIMO communication," \emph{arXiv preprint}, arXiv:1903.08127, 2019.


\bibitem{Pan2019}
C. Pan, H. Ren, K. Wang, M. Elkashlan, A. Nallanathan, J. Wang, and L. Hanzo, ``Intelligent reflecting surface aided
mimo broadcasting for simultaneous wireless information and power transfer," \emph{arXiv preprint}, arXiv:1908.04863, 2019.

\bibitem{Basar2019}
E. Basar, ``Reconfigurable Intelligent Surface-Based Index Modulation: A New Beyond MIMO Paradigm for 6G," \emph{arXiv preprint}, arXiv:1904.06704, 2019.


\bibitem{Fu2019}
M. Fu, Y. Zhou, and Y. Shi, ¡°Intelligent reflecting surface for downlink non-orthogonal multiple access networks,¡± \emph{arXiv preprint}, arXiv:1906.09434, 2019.

\bibitem{ZLI2019}
Z. Li, M. Hua, Q. Wang, Q. Song, `` Weighted Sum-Rate Maximization for Multi-IRS Aided Cooperative Transmission," \emph{arXiv preprint}, arXiv: 2002.04900, 2020.



\bibitem{Wang2010}
X. Wang, and X. Zhang, ``Optimal Beamforming in MIMO Two-Way Relay
Channels,'' in \emph{Proc. IEEE Global Telecommun. Conf.}, pp. 1-5,
Dec. 2010.

\bibitem{Zhang2016}
Z. Zhang, Z. Chen, M. Shen, and B. Xia, ``Spectral and energy efficiency of multipair two-way full-duplex relay systems
with massive MIMO," \emph{IEEE J. Select. Areas Commun.}, vol. 34, no. 4, pp. 848-863, 2016.


\bibitem{Blahut1972}
R. Blahut, ``Computation of channel capacity and rate-distortion functions,"
\emph{IEEE Trans. Inf. Theory}, vol. 18, no. 4, pp. 460-473, July 1972.

\bibitem{Wang2010TWC}
X. Wang and X. D. Zhang, ¡°Linear transmission for rate optimization in
MIMO broadcast channels,¡± \emph{IEEE Trans. Wireless Commun.}, vol. 9, no. 10, pp. 3247-3257, 2010.
%
\bibitem{Cover1991}
T. M. Cover and J. A. Thomas, Elements of Information Theory. New
York: John Wiley \& Sons, Inc., 1991.
%
\bibitem{Kay1993}
S. M. Kay, Fundamentals of Statistical Signal Processing: Estimation
Theory. Upper Saddle River, NJ: Prentice Hall PTR, 1993.
%

\bibitem{Boyd2004}
S Boyd, L Vandenberghe, Convex Optimization, Cambridge University Press, 2004.
%

\end{thebibliography}
\end{document}